# Global Major-Element Maps of Mercury from Four Years of MESSENGER X-Ray Spectrometer Observations

Larry R. Nittler[1], Elizabeth A. Frank[1], Shoshana Z. Weider[1], Ellen Crapster-Pregont[2,3], Audrey Vorburger[2], Richard D. Starr[4,5], and Sean C. Solomon[3].

[1]Earth and Planets Laboratory, Carnegie Institution for Science, Washington, DC 20015, USA.

[2]Department of Earth and Planetary Sciences, American Museum of Natural History, New York, NY 10024, USA.

[3]Lamont-Doherty Earth Observatory, Columbia University, Palisades, NY 10964, USA.

[4]Physics Department, The Catholic University of America, Washington, DC 20064, USA.

[5]Solar System Exploration Division, NASA Goddard Space Flight Center, Greenbelt, MD 20771, USA.

ABSTRACT

The X-Ray Spectrometer (XRS) on the MESSENGER spacecraft provided measurements of major-element ratios across Mercury's surface. We present global maps of Mg/Si, Al/Si, S/Si, Ca/Si, and Fe/Si derived from XRS data collected throughout MESSENGER's orbital mission. We describe the procedures we used to select and filter data and to combine them to make the final maps, which are archived in NASA's Planetary Data System. Areal coverage is variable for the different element-ratio maps, with 100% coverage for Mg/Si and Al/Si, but only 18% coverage for Fe/Si north of 30° N, where the spatial resolution is highest. The spatial resolution is improved over previous maps by 10–15% because of the inclusion of higher-resolution data from late in the mission when the spacecraft periapsis altitude was low. Unlike typical planetary data maps, however, the spatial resolution of the XRS maps can vary from pixel to pixel, and thus care must be taken in interpreting small-scale features. We provide several examples of how the XRS maps can be used to investigate elemental variations in the context of geological features on Mercury, which range in size from single ~100-km-diameter craters to large impact basins. We expect that these maps will provide the basis for and/or contribute to studies of Mercury's origin and geological history for many years to come.

1. Introduction

Prior to the MErcury Surface, Space ENvironment, GEochemistry, and Ranging (MESSENGER) mission (Solomon et al., 2007), most of our knowledge about Mercury was drawn from the three Mariner 10 flybys in 1974 and 1975. Mariner 10's observations of the innermost planet's surface, however, were limited: there was no geochemical remote sensing instrument on board, and because of the time interval between flybys, only 40–45% of Mercury's surface was imaged.

More than 30 years later, MESSENGER's X-Ray Spectrometer (XRS; Schlemm et al., 2007) and Gamma-Ray and Neutron Spectrometer (GRNS; Goldsten et al., 2007) returned the first compositional measurements of Mercury's surface (Nittler et al., 2011; Peplowski et al., 2011). These observations revealed, for the first time, that Mercury's crust is rich in Mg and S, but poor in Al, Ca, and Fe relative to terrestrial and lunar materials (Evans et al., 2012; Nittler et al., 2011). The crust is also abundant in K (Peplowski et al., 2011, 2012), Na (Peplowski et al., 2014), and Cl (Evans et al., 2015). Petrological experiments and modeling based on early MESSENGER compositional results indicated that Mercury's surface, on average, is most similar to magnesian basalts (Stockstill-Cahill et al., 2012) and Fe-poor basaltic komatiites (Charlier et al., 2013).

Observations made during the first year of MESSENGER orbital observations documented the chemical heterogeneity of Mercury's surface—though muted in comparison with Earth and the Moon—and the presence of a compositional dichotomy (Peplowski et al., 2012; Weider et al., 2012). As the periapsis altitude of MESSENGER's eccentric, near-polar orbit declined during the last two years of orbital operations, the spatial resolution of XRS observations in the northern hemisphere of Mercury significantly improved. With these later, higher-resolution measurements, the presence of several geochemically distinct regions, dubbed "geochemical terranes," was revealed (Weider et al., 2015; hereafter W15) and independently confirmed by Neutron

Spectrometer (NS) measurements of neutron absorption (Peplowski et al., 2015). The existence of the geochemical terranes most likely reflects variable degrees of partial melting of a chemically heterogeneous mantle. The derived terrane compositions have subsequently been used for petrologic experiments and modeling to investigate the geochemical history of Mercury's crust (McCoy et al., 2018; Peplowski and Stockstill-Cahill, 2019; Vander Kaaden and McCubbin, 2016; Vander Kaaden et al., 2017). However, different authors have used different criteria (often a mixture of arbitrary element abundance thresholds and geomorphological unit boundaries) to define terranes, and as yet there is no overall consensus on the definition or number of such terranes.

The first identification of geochemical terranes from XRS observations was based on data that were obtained through December 2013 (W15). During that mission phase, the XRS spatial resolution was limited to >200 km. During the final year of the mission, however, because MESSENGER regularly reached periapsis altitudes less than 200 km, the highest-resolution measurements in the course of the mission (as good as several kilometers for some late-mission XRS data) were obtained. Major-element ratio maps of Mercury derived from a comprehensive analysis of the full MESSENGER orbital dataset have been available in NASA's Planetary Data System (PDS) since 2016. They have already been used as the basis for several investigations into Mercury's origin and geological evolution (Frank et al., 2017; McCoy et al., 2018; Nittler et al., 2018; Vander Kaaden et al., 2017). In this paper, we provide a detailed account of how these final maps were produced, compare the final maps with previously published maps, and discuss a few examples of specific regions on Mercury where observed elemental variations may provide new insights into the planet's geological history.

## 2. MESSENGER'S X-Ray Spectrometer

MESSENGER's XRS measured the surface abundances, by weight, of the rock-forming elements Mg, Al, Si, S, Ca, Ti, and Fe (Schlemm et al., 2007). The measurements were made via planetary X-ray fluorescence (XRF), which has been widely used to determine remotely the surface composition of airless, rocky planetary bodies in the inner solar system (Prettyman et al., 2019). This technique is reliant on X-rays emitted from the Sun's corona that excite electrons in atoms on or very near the surface (within the top few tens of micrometers) of the body. Upon the return of an atom to a stable energetic state, a fluorescent X-ray, with characteristic energy indicative of the element, is emitted. By simultaneously measuring the solar X-ray spectrum (which is highly variable in time, in terms of absolute flux and spectral shape) and the spectrum of X-ray fluorescence from the planet, abundance ratios of detected elements can be determined to reveal the chemical abundances on the planet's surface. During typical "quiet-Sun" conditions, when coronal plasma temperatures are just a few megakelvin (MK), only fluorescent Mg, Al, and Si are detectable. In contrast, during solar flares the coronal plasma temperature can reach tens of MK, and heavier elements (S, Ca, Ti, Fe) also fluoresce (Nittler et al., 2011).

The MESSENGER XRS consisted of three gas-proportional counter (GPC) detectors that tracked the planet, and a Sun-pointed Si-PIN detector in the Solar Assembly for X-Rays (SAX) (Schlemm et al., 2007; Starr et al., 2016), all of which had an energy range of 1–10 keV. As the energy resolution of the GPCs was relatively poor, a "balanced filter" method (Adler et al., 1972) was used to separate the adjacent fluorescent $K_\alpha$ lines of Mg, Al, and Si (at 1.25, 1.49, and 1.74 keV, respectively). Thin foils of Mg and Al covered two of the GPC detectors to selectively absorb X-rays at different energies. This approach permits Mg, Al, and Si signals to be distinguished in XRF spectra (Adler et al., 1972; Trombka et al., 2000). This detection challenge did not exist for higher-atomic-number elements because the energy resolution of the detectors was sufficient to separate the higher-energy characteristic fluorescent lines via spectral fitting. While in orbit, MESSENGER regularly encountered charged particles, e.g., ions and energetic electrons in Mercury's magnetosphere (Korth et al., 2018). These particles occasionally caused enhanced backgrounds in the XRS detectors as well as fluorescence of the collimators and filter foils (Slavin et al., 2008; Weider et al., 2014), in many cases rendering spectra unusable for geochemical analysis.

The collimators on the GPCs had a hexagonal 12° field of view (FOV), meaning that the planet-facing detector measurements had a footprint derived from a projection of this hexagonal FOV onto the planetary surface—stretched in the direction of spacecraft motion (Fig. S1). Each XRS spectrum thus reflects the average composition of the surface within its respective footprint, weighted by the non-uniform response of the collimator (Starr et al., 2016). Specific footprint sizes and integration times for any particular XRS measurement were dictated by MESSENGER's distance from Mercury. Given the spacecraft's highly eccentric orbit, these distances varied greatly from periapsis to apoapsis, ranging from >3000 km in effective diameter for the southern hemisphere to a few kilometers in the north during the later parts of the mission (Starr et al., 2016;

Weider et al., 2012).

**3. Methods and data**

The data selection criteria and analysis methods we used in this work were almost identical to those described by W15, with a few important modifications discussed in the following sections. All our XRS analysis was carried out with custom software written in the IDL programming language (Harris Geospatial Solutions) and used routines from the SOLARSOFT solar physics package (Bentely and Freeland, 1998). In addition, we used ArcGIS (Esri) to overlay the XRS maps with other MESSENGER datasets.

To date, elemental abundances and maps derived from the XRS data have all been normalized to the abundance of Si, as using ratios eliminates some systematic uncertainties and Si is a major element for which the surface abundance generally varies less (<15%) than those of other major elements (Peplowski et al., 2012; Weider et al., 2012; W15). Likewise, the maps we report here are presented as elemental ratios (to Si). The Mg/Si and Al/Si maps have global coverage, as these elements can be measured during quiet-Sun periods. In contrast, even after four years of orbital data collection, the coverage for S/Si, Ca/Si, and Fe/Si maps is not global. This is because solar flares—which occur only sporadically—were required to measure these heavier elements.

*3.1. Quiet-Sun data*

Our Mg/Si and Al/Si maps include data from both quiet-Sun and solar-flare periods from throughout MESSENGER's four-year orbital mission. For the quiet-Sun analysis, we initially selected all spectral integration periods during which the XRS FOV included sunlit Mercury and the solar plasma temperature inferred from the SAX solar monitor was <10 MK. However, because the southern hemisphere is greatly oversampled in the XRS dataset, due to the spacecraft's eccentric orbit, we cut the dataset by including only a random sampling of the southern-hemisphere XRS integrations, whereas all footprints in the northern hemisphere were considered. Although southern footprints made up >50% of all XRS measurements, about 80% of the final quiet-Sun data used to generate the element maps were from the northern hemisphere. We further reduced the size of the quiet-Sun dataset used for our final maps, before and after their analysis, following the criteria used by W15 (i.e., integrations with spectra indistinguishable from the typical detector background and those that exhibited charged-particle contamination, and from northern-hemisphere footprints that were larger than average for that latitude). In the end, data from 46,663 individual quiet-Sun XRS measurements (Table 1) were included in the Mg/Si and Al/Si maps presented here—about twice the number used to generate the maps of W15.

As described by W15, the quiet-Sun data needed to be background-subtracted, binned, and inverted (see Eq. (1) of W15) to derive the Mg, Al, and Si XRF fluxes incident on the instrument. The binning procedure of W15 was developed because of the low signal-to-noise ratio of individual quiet-Sun integrations that often leads to very large errors on derived Al/Si ratios (the signal-to-noise ratios for Mg/Si are better because of Mercury's higher Mg concentration). In the spatial binning procedure of W15, Mercury's surface was divided into 3°×3° latitude–longitude bins, and all quiet-Sun XRS measurements with a footprint center within a given bin were summed and analyzed as a single integration. This method has the advantage of improving signal-to-noise ratio, but at the expense of spatial resolution.

We modified the binning procedure for the production of our final maps to take advantage of the larger data set and the higher-resolution measurements obtained during the last year of MESSENGER's orbital mission. Although XRS footprints are hexagonal in shape and often stretched out in the north–south direction by the spacecraft motion, we calculated an effective footprint diameter for each spectral integration from the diameter of a circle with the same area. Whereas W15 used spatial bins of varying size, here we generated equal-area bins to optimize spatial resolution. Measurements with footprint diameters larger than 100 km were binned into ~66×66 $km^2$ bins, those from footprints 50–100 km across were binned into ~26×26 $km^2$ bins, and those from footprints smaller than ~50 km in diameter were not binned at all. This procedure favors spatial resolution over statistical precision for the highest-resolution data. The number of spectra included for each size range is shown in Table 1. This binning procedure reduced the total number of quiet-Sun spectra that we used to determine Mg/Si and Al/Si ratios from 46,663 to 22,801.

To convert the Mg/Si and Al/Si photon ratios obtained from the binning procedure to element abundance ratios, we used theoretically derived calibration curves that relate the measured photon ratios to solar plasma temperature (see Fig. S2 and Fig. 2a of W15), and we empirically corrected these curves to match the more-accurate solar-flare data. To infer the solar plasma temperatures for each quiet-Sun integration, we used Eq. (2) of W15 (a derived quadratic relationship that accounts for the shape of the solar spectrum, as measured by the SAX solar monitor). A list of all 46,663 quiet-Sun spectra used in the maps along with corresponding binning information and Mg/Si and Al/Si ratios is provided as supplementary information.

*3.2. Solar flare data*

In this work, we used the previously described forward-modeling method (Nittler et al., 2011; Weider et al., 2012, 2014, 2016; W15) to analyze the XRS data acquired during solar flares. In total, we fit nearly 2300 XRS solar-flare spectra—acquired during 291 distinct solar flares—including fits to individual integrations during flares and to spectra summed over entire flare periods. Prior to generation of our final maps, however, we reduced the magnitude of the flare dataset on the basis of the W15 criteria (this mainly involved removing spectra with obvious charged-particle contributions and spectra for which derived abundances differed greatly from multiple overlapping measurements). Our maps for the different element ratios include variable numbers of flare analyses (see Table 1) because stronger solar flares (which occur less frequently) are required to derive abundances for the heavier elements such as Ca and (especially) Fe than for the lighter elements (i.e., Mg, Al, and Si). In addition, we required relatively fewer flare data for the Mg/Si and Al/Si maps because of the large number of quiet-Sun data available for those elements. A spreadsheet containing the fitting results for all flare spectra used in generating the maps is provided as supplemental material.

*3.2.1. The case of Fe*

It was possible to measure Fe with the XRS during only the most powerful solar flares (when the solar X-ray flux is much stronger, especially at higher energies) because its primary $K_\alpha$ X-ray line is at substantially higher energy than those of the other elements discussed here. In addition, as discussed in detail by Weider et al. (2014), the accurate determination of Fe abundances from MESSENGER XRS data presents distinct challenges compared with the other major elements. Of particular importance are an observed dependence of Fe/Si ratios on phase angle, as well as the possible effects of Fe segregation into specific minerals.

The fundamental-parameters modeling approach used to convert XRS spectral data into elemental abundances (Clark and Trombka, 1997; Nittler et al., 2011) is based on the assumption that the analyzed sample (i.e., Mercury's surface) is flat and chemically homogeneous. At large phase angles, shadowing effects from surface topography can cause deviations in XRF fluxes from those predicted by the fundamental-parameters modeling (Maruyama et al., 2008; Naranen et al., 2008; Okada, 2004; Weider et al., 2011). Weider et al. (2014) observed—from 55 flare spectra XRS measurements—a correlation between derived Fe/Si ratios and phase angle, and from this correlation they developed an empirical correction. The larger dataset used here confirms that correlation (Fig. S3). Indeed, a linear fit to Fe/Si as a function of phase angle for 117 flare measurements with large footprints in Mercury's southern hemisphere (in our data set) yields Fe/Si = 0.00191 $\phi$ – 0.11068, where $\phi$ is the phase angle in degrees (solid line in Fig. S3a)—a relation that is slightly different from the one derived by Weider et al. (2014). We removed the phase-angle dependence from the flare measurements by dividing the measured Fe/Si ratio by the ratio predicted for its phase angle from this linear fit, and then renormalized the data set such that the average Fe/Si value was equal to that of the uncorrected data (0.06). The resulting corrected data for all 262 flare measurements included here are shown in Fig. S3. No significant correlation between elemental ratio and phase angle was observed for the other mapped elements (Fig. S4).

Iron on Mercury's surface is highly reduced and most likely largely incorporated into metal and/or sulfide phases in the regolith (Weider et al., 2014). If such grains are smaller than a few tens of micrometers (the relevant depth scale for production of XRF) and finely mixed with silicates and other phases, the assumption within the fundamental-parameters methodology of a homogeneous sample is probably a reasonable one. If, however, the Fe-rich grains are larger, this assumption could break down and the expected XRF fluxes could differ strongly from the model predictions. Weider et al. (2014) modeled this "mineral mixing" effect and found that it could change the observed Fe/Si X-ray photon ratios from Mercury's surface by up to a factor of five. Without detailed knowledge of the sizes, shapes, and chemical compositions of regolith grains on Mercury, it is not possible to estimate accurately the magnitude of any required correction to the ratios, and thus there is a very large systematic uncertainty to the overall normalization of Fe/Si from the XRS data. As discussed above, we applied an arbitrary normalization to the phase-angle-corrected data such that the average value of the corrected Fe/Si values is 0.06. This figure is slightly lower than the average northern hemisphere Fe/Si ratio of 0.077±0.013, derived from MESSENGER gamma-ray data (Evans et al., 2012), suggesting that the overall XRS values are underestimated, but perhaps not by as much as the mineral mixing model indicates. In any case, some spatial variations observed in the Fe/Si map are correlated with other elemental variations across Mercury's surface, so even if the overall normalization is uncertain, the map reflects real heterogeneity in the distribution of Fe.

*3.3 Map generation*

Generating elemental maps from MESSENGER XRS data requires combining measurements with variable spatial resolution and uncertainty. Here, we used the method developed in previous work (Weider et al., 2014; W15). Briefly, we divided Mercury's surface into 0.25°×0.25° pixels in cylindrical projection. For each pixel, we identified all XRS footprints in the dataset (flare and quiet-Sun) that overlap with it and calculated a weighted average for each element ratio, $R$:

$$R_{\text{av}} = \sum_{i=1}^{N} w_i R_i \ / \ \sum_{i=1}^{N} w_i.$$

In this equation, $R_i$ is the value of $R$ in the $i^{\text{th}}$ of a total of $N$ overlapping footprints, and $w_i$ is a weighting factor: $w_i = 1/(A_i \times \sigma_i^2)$, where $A_i$ is the area of footprint $i$ and $\sigma_i$ is the uncertainty in $R_i$. The uncertainty is taken as the larger of either the statistical uncertainty returned by the fitting procedure or an assumed minimum relative systematic uncertainty of 5% (10% for Fe/Si). The weighting procedure thus favors overlapping footprints that have the smallest areas and/or smallest analytical uncertainty. For each elemental ratio we generated several additional maps (provided as supplemental material), with each using the same weighting factors as the elemental data. These included the following: (i) a statistical-error map (see Eq. (5) of W15); (ii) an effective resolution map, where resolution is defined as the diameter of a circular region with area equal to the weighted average of $A_i$ values for individual footprints; and (iii) a footprint asymmetry map, where each individual footprint is assigned an asymmetry value (Fig. S1) based on the maximum north–south extent of the footprint divided by the maximum east–west extent, and the weighted average of these was found for each pixel (this map is useful for smoothing, as discussed below).

The boundaries of the XRS footprints themselves were calculated from MESSENGER navigational data and are provided (in the PDS) as a set of latitude–longitude pairs that define the perimeter of the full FOV of the XRS planet-facing detectors for each spectral integration in which the FOV included a portion of Mercury's surface. Note, however, that the XRS response is close to Gaussian in shape across the collimator, with a full-width-at-half-maximum of ~6°, compared with the full 12° instrument FOV (Starr et al., 2016). Moreover, the solar X-ray illumination did not always uniformly fill the FOV because of surface topography and shadowing effects. Therefore, the true spatial resolution of the individual XRS measurements is generally better than that derived here from the footprint files. This difference is taken (approximately) into account by our smoothing procedure, as discussed further below. In principle, more accurate and higher-resolution maps could be generated by considering, in detail, both the collimator response and the solar illumination conditions for each measurement, but doing so would require a far more computationally complex approach than the steps described here.

Our basic map-generation method results in some obvious image artifacts, along which sharp borders of individual footprints are sometimes clearly present (Fig. 1a). As was done by W15, we thus performed an additional smoothing step in which the value of each pixel in a map was replaced by the geometric mean value of all pixels in a region surrounding it. The corresponding uncertainty in each pixel was taken to be the standard deviation of values in the smoothing region (or the original pixel uncertainty if all pixels within the smoothing region originate from a single XRS measurement). The smoothing region adopted for a given pixel by W15 was circular (with a diameter equal to the effective resolution-map value at that pixel). In contrast, we used elliptical smoothing regions, each with an area 50% that of the weighted average footprint area for a given pixel, stretched out in either the north–south or east–west direction according to the footprint asymmetry map described above (see Fig. S5). This smaller area takes into account the non-uniform collimator response and solar illumination issues discussed above. Furthermore, the elliptical shape takes into account the fact that many XRS footprints—because of MESSENGER's near-polar orbit— are elongated along track in the north–south direction (east–west at very high latitudes), especially for the high-resolution measurements obtained at low altitudes during the last year of the mission. Our smoothing process, therefore, generally results in better resolution in the cross-track direction in northern regions where low-altitude data were obtained. In Fig. 1 we compare the unsmoothed (Fig. 1a) map of Mg/Si to the smoothed version (Fig. 1b) for a region of Mercury's northern hemisphere. (Note that we have not utilized the rainbow color ramp adopted in previous XRS publications.) It is important to recognize that—as a result of the way they are generated—the XRS maps differ from many other types of mapped planetary data, i.e., because each pixel value represents not just that point on the surface, but the average over an area surrounding it, and because the size of the area (resolution) varies from pixel to pixel. Thus, what appear in some places to be sharp compositional borders are instead often representations of abrupt changes in resolution. This issue is discussed in more detail below.

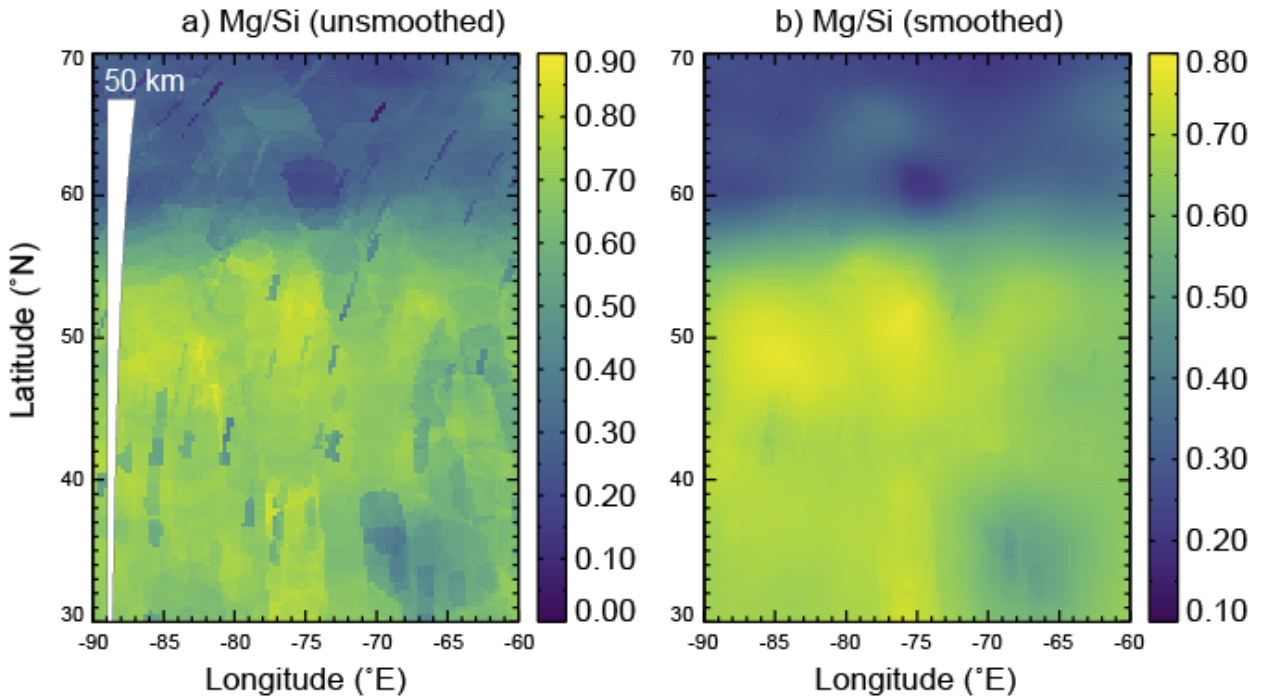

**Fig. 1.** (a) Unsmoothed and (b) smoothed maps of Mg/Si for a region of Mercury's northern hemisphere. The location of this region and others highlighted throughout this paper can be seen on a monochrome Mercury Dual Imaging System (MDIS) global image mosaic in Fig. S6. Note that the smoothing process reduces the total dynamic range of the map, accounting for the narrower data range in (b).

## 4. Results

All maps (unsmoothed, smoothed, errors, spatial resolution, and footprint asymmetry) are provided in binary format as supplementary material.

*4.1 Mg/Si and Al/Si maps*

The smoothed maps of Mg/Si and Al/Si are shown in Fig. 2, along with their associated uncertainty maps, a map of Al/Mg derived from them, and a map of effective spatial resolution. Because these maps were derived from both quiet-Sun and solar flare data, they have full global coverage. The highest spatial resolution of these maps occurs at mid-to-high northern latitudes, where it is better than 200 km on average and is as small as 20 km in some places. However, as discussed above, these are upper limits on the true resolution. The broad-scale chemical heterogeneity across Mercury's surface that has been identified in previous work—and particularly the presence of distinctive geochemical terranes—is still clearly apparent, but the higher spatial resolution of these new maps also reveals additional details that we explore further below.

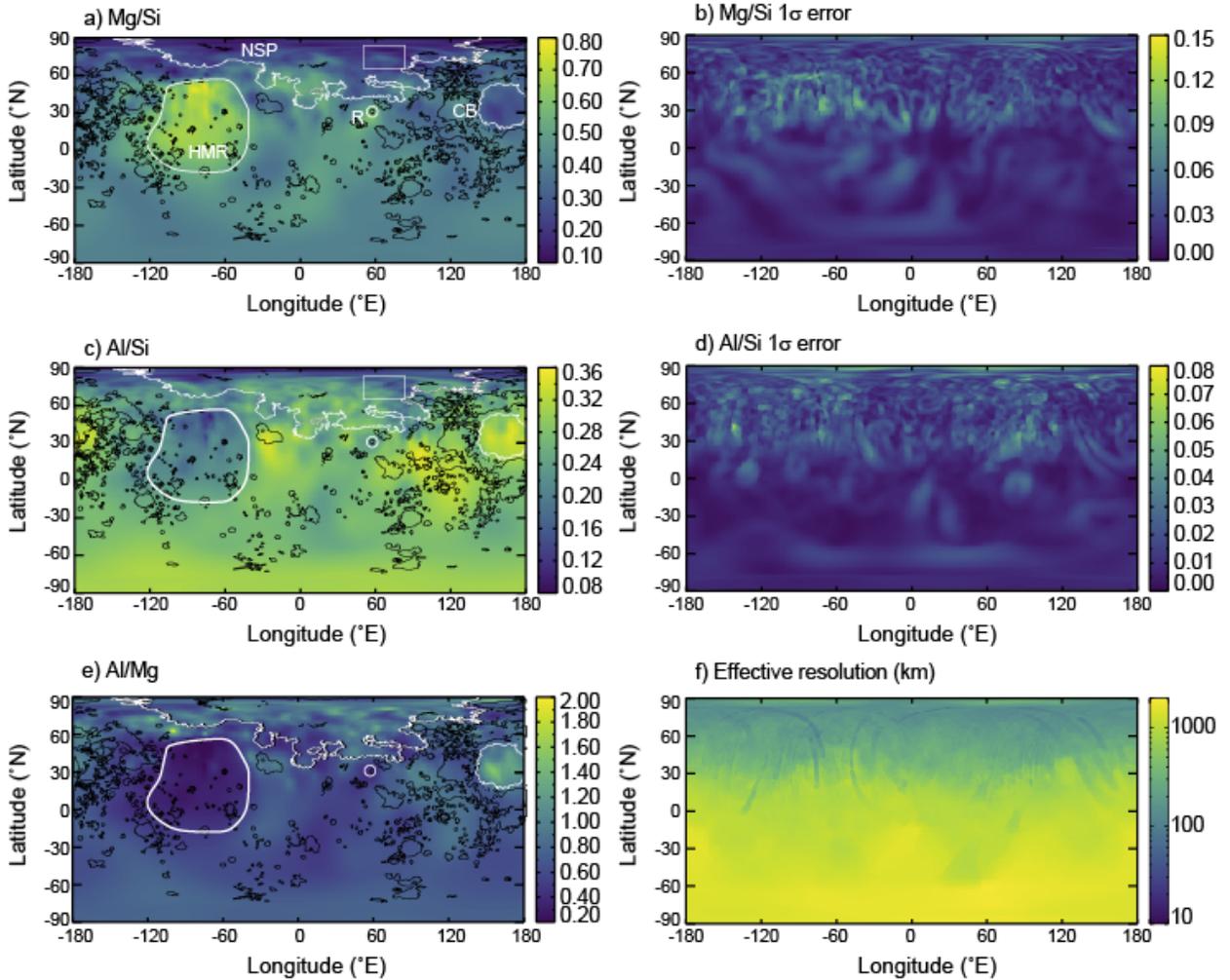

**Fig. 2.** Smoothed maps of (a) Mg/Si, (c) Al/Si, and (e) Al/Mg, as well as (b, d) their associated uncertainties (where σ is the standard deviation) and (f) effective resolution, all in cylindrical projection. Some major features are indicated by white outlines: CB = Caloris basin; NSP = northern smooth plains (Head et al., 2011); HMR = high-Mg region (Weider et al., 2015); R = Rachmaninoff basin. Smooth plains deposits (Denevi et al., 2013) are outlined in black. The white rectangle in panels (a) and (c) indicates the location of the impact craters in the NSP discussed in Section 5.2.3.

*4.2 S/Si and Ca/Si maps*

The smoothed maps of S/Si and Ca/Si are shown in Fig. 3 along with their associated uncertainties, as well as a map of Ca/S derived from them. Because these elements required solar flares for their detection, spatial coverage is incomplete in the northern hemisphere, although considerably improved over previously published maps that were produced from less than three years of data (W15). The most striking feature of the S/Si and Ca/Si maps is the obvious enhancement of both S and Ca in the high-Mg region. Since the first MESSENGER orbital data were available (e.g., Nittler et al., 2011), it has been recognized that S and Ca are strongly correlated on Mercury's surface, suggestive of the presence of oldhamite (CaS). This correlation is reflected in our Ca/S map, which shows an almost uniform Ca/S ratio across the planet. The marked exception is the ~75-km-diameter pyroclastic deposit (Nathair Facula) northeast of the Rachmaninoff basin. The unusually low S and C abundances (the latter determined from MESSENGER Neutron Spectroscopy measurements; Peplowski et al., 2016) in this area have been interpreted as evidence that the Nathair Facula eruption was driven by exsolution of S- and C-bearing volatile species (Weider et al., 2016).

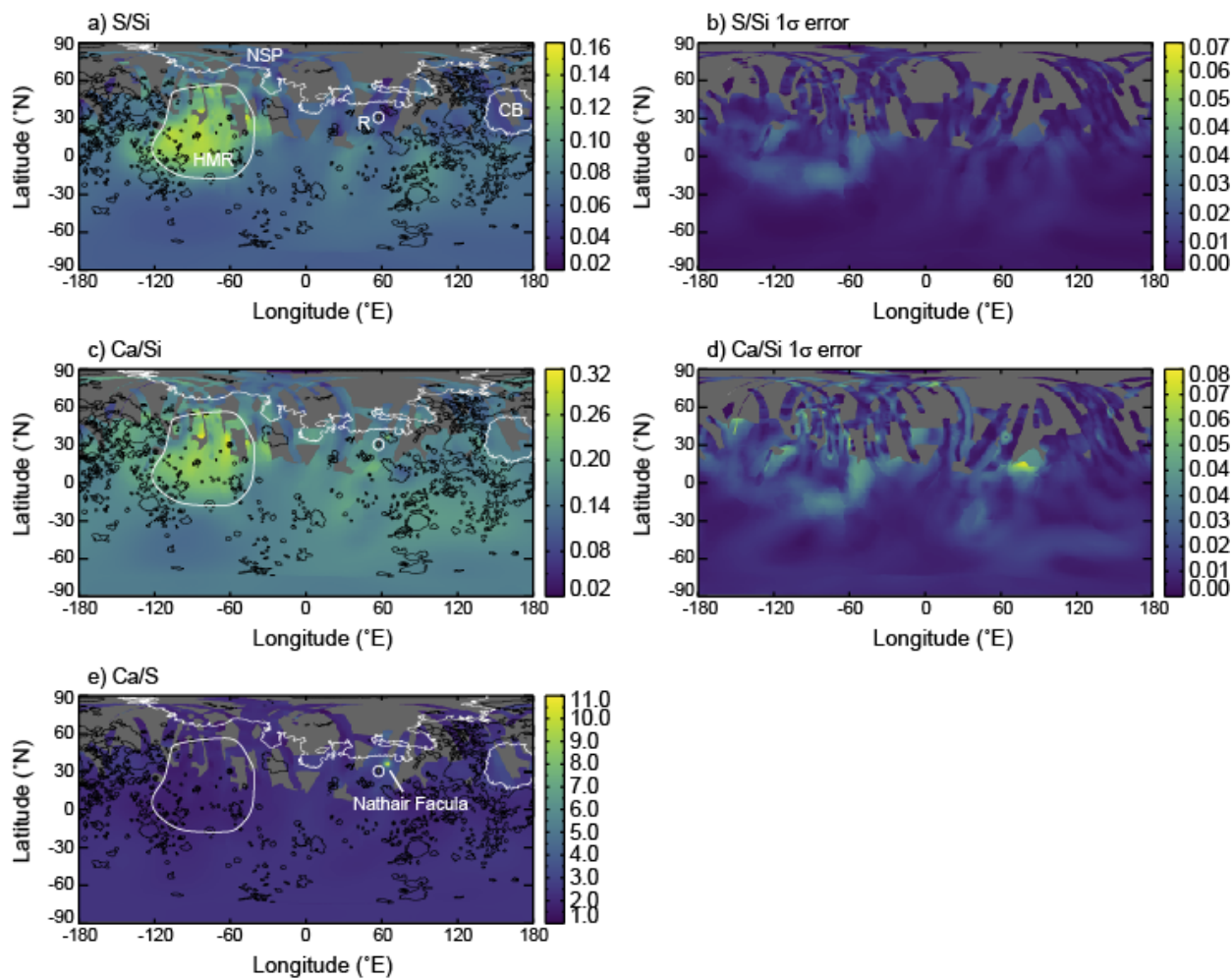

**Fig. 3.** Smoothed maps of (a) S/Si, (c) Ca/Si, and (e) Ca/S, as well as (b, d) their associated uncertainties, all in cylindrical projection. Some major features are indicated by white outlines and smooth plains deposits are outlined in black (as in Fig. 2): The location of the large pyroclastic deposit (Nathair Facula) northeast of Rachmaninoff has an unusually high Ca/S ratio, as indicated in (e). Grey areas are unmapped for these elements.

*4.3 Fe/Si map*

The smoothed Fe/Si map and its associated uncertainty map are shown in Fig. 4. Even with more than four years of orbital data, Fe/Si data cover only about 40% of the northern hemisphere, and only 18% of the area northward of 30° N, where the spatial resolution is highest. Nevertheless, clear variations in Fe/Si across both the northern and southern hemispheres are observed. In some cases, these variations are correlated with variations in other elements. For example, the high-Mg region contains some of the highest Fe/Si ratios observed, and a region to its southeast is notably lower than average for both Mg/Si and Fe/Si, as are the Caloris basin interior plains. On the other hand, there is an apparent region of relatively high Fe/Si near 60° E longitude and the equator that does not stand out in any of the other element-ratio maps.

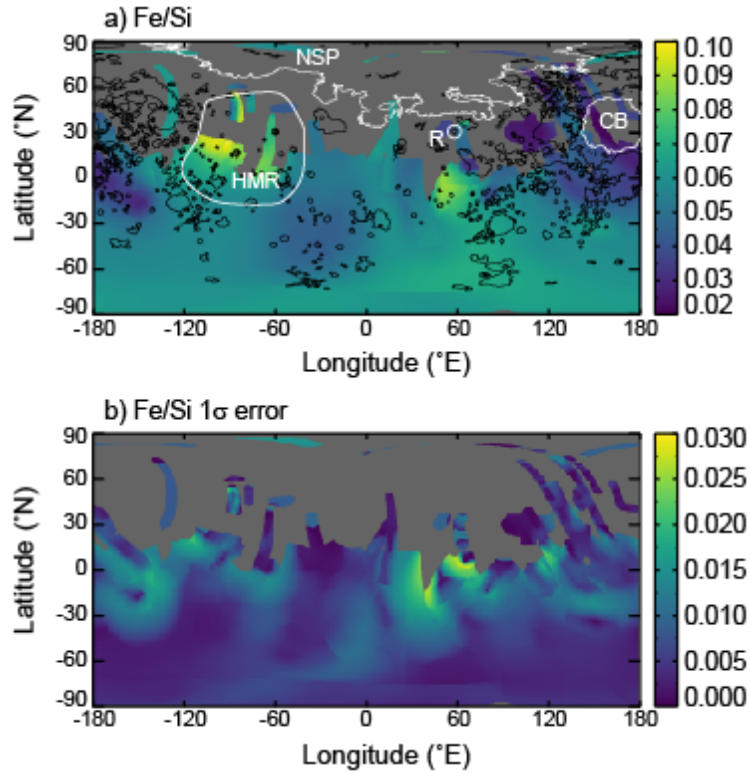

**Fig. 4.** Smoothed maps of (a) Fe/Si and (b) its associated uncertainty, both in cylindrical projection. Major features and smooth plains deposits are outlined in white and black, respectively (as in Fig. 2 and 3). Grey areas are unmapped for Fe/Si.

## 5. Discussion

### 5.1. Improvement in spatial resolution

Our new elemental ratio maps of Mercury—specifically the Mg/Si and Al/Si maps with global coverage—have spatial resolution superior to those published previously by W15. This improvement is illustrated in Fig. 5, which compares profiles of Mg/Si and Al/Si—from the old (W15) and new maps—across the ~105-km-diameter impact crater Akutagawa. The profiles through the previous and new effective-resolution maps (bottom panel of Fig. 5c) show that, on average, the resolution of the new maps is 15–20% better and that some pixels show an even greater improvement. Note that the effective-resolution profiles have been scaled downward from the map values by a factor of $1/\sqrt{2}$ to take into partial account the overestimation of illuminated footprint size (Section 3), but even these are upper limits and the true resolution is likely better.

In the new maps, Akutagawa clearly shows enhanced Mg/Si and lower Al/Si than the surrounding material. In addition, an ~25-km-wide feature with even higher Mg and lower Al, and apparently sharp compositional changes, can be seen between ~400 and 425 km along the profile. The corresponding resolution profile, however, illustrates that the boundaries of this feature are not caused by sharp compositional changes, but rather are an indication that these pixels have higher spatial resolution than the neighboring sections of the profile. To illustrate this point further, the smoothing regions used for two adjacent pixels in the profile (vertical lines in Fig. 5c) are indicated as ellipses in Fig. 5b; these ellipses represent the areas that were averaged to give the map values (see Section 3.3). Pixel 2 is centered on the crater and has better spatial resolution (smaller smoothing region) than pixel 1. Pixel 1 includes more material from the area surrounding the crater in its average. The apparent 25-km width of the Mg/Si peak is therefore misleading, as the feature resolved by this peak is actually the 100-km-diameter crater. This result highlights one of the challenges associated with the interpretation of maps that have highly variable spatial resolution, and we note that compositional profiles derived from these maps should be interpreted with care.

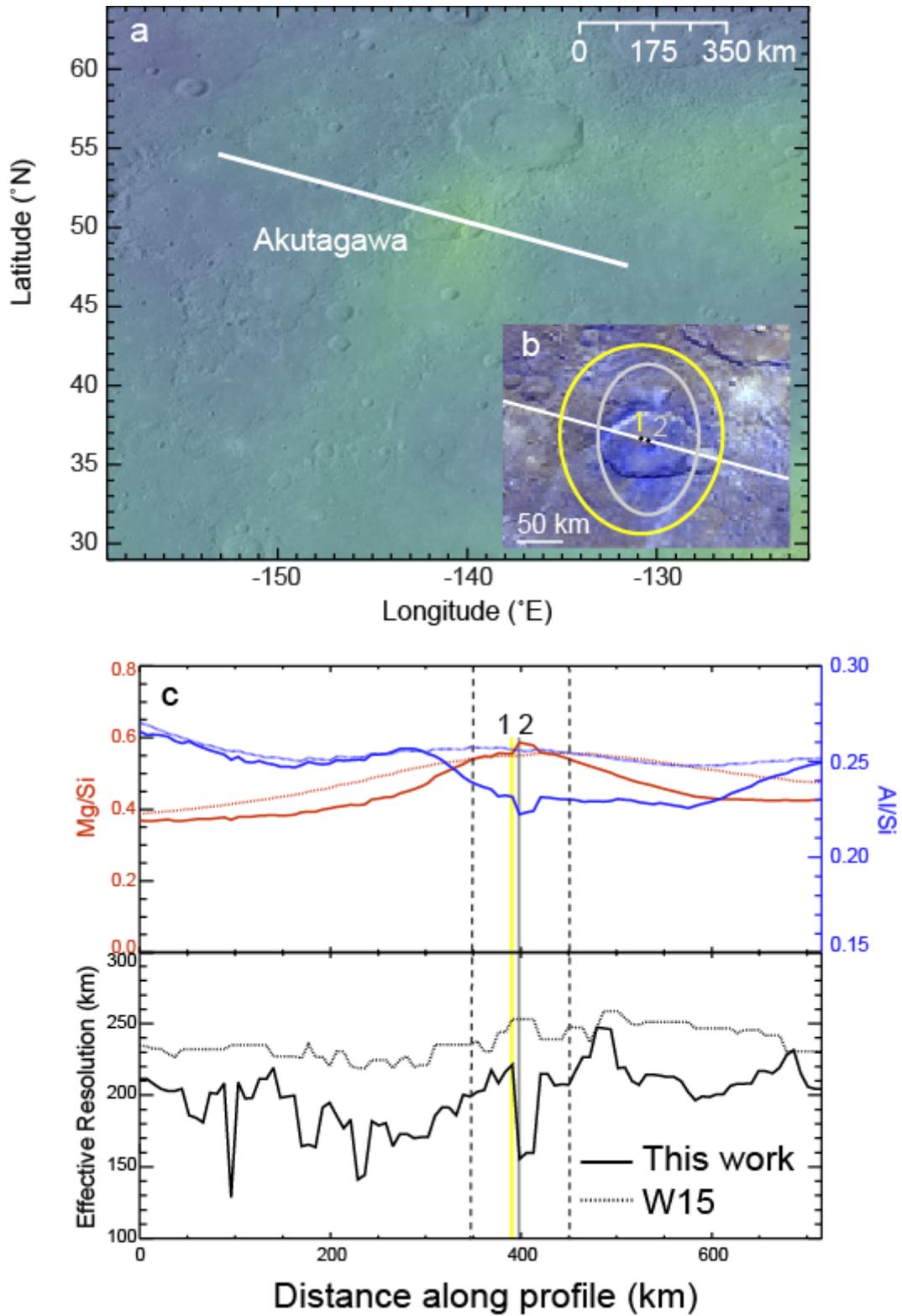

**Fig. 5.** (a) False-color map of Mg/Si (color scale is the same as in Fig. 2a) overlain on a Mercury image mosaic, derived from MDIS observations, centered near Akutagawa crater at 48°N and –141°E. The white line indicates the location of the profiles shown in (c). (b) Enhanced-color MDIS

image of the area around Akutagawa; in this representation (Robinson et al., 2008; Denevi et al., 2009) low-reflectance material appears blue (see Section 5.2.5). The ellipses indicate smoothing regions for the two indicated pixels along the profile line. (c) Profiles of Mg/Si and Al/Si and effective resolution (scaled by $1/\sqrt{2}$ from map values to correct partially for overestimation of footprint size) along the path indicated by the white line in (a). Solid profiles in (c) are derived from the maps reported here; dotted profiles are from the maps of W15. Vertical yellow and grey lines indicate the locations along the profiles of the two pixels indicated in (b), and vertical dashed lines denote the edges of Akutagawa.

Akutagawa is highly enriched in low-reflectance material (LRM; blue in Fig. 5b), which generally appears on Mercury's surface in deposits excavated from deeper in the crust by impact cratering events (Ernst et al., 2010). In addition, an enhanced flux of thermal neutrons was measured over this crater, which indicates the presence of several wt% C (Klima et al., 2018; Peplowski et al., 2016). The high C contents associated with LRM have been interpreted to be the remnants of a primary graphite floatation crust formed from an early magma ocean on Mercury (Vander Kaaden and McCubbin, 2015). The high-resolution XRS measurement described here, the first for an LRM deposit, clearly shows that this C-enriched LRM is also enriched in Mg and depleted in Al relative to brighter material surrounding the crater. This finding, and the Mg/Si and Al/Si values for Akutagawa, are consistent with earlier reports for the composition of the Rachmaninoff basin (W15), which contains a highly concentrated LRM deposit (e.g., Klima et al., 2018). Below we discuss additional specific examples of geological and/or geochemical features for which the new higher-resolution maps reveal geochemical aspects that were not readily evident in the earlier maps.

*5.2. A closer look at some specific features*
*5.2.1. High-Mg region*

Perhaps the most conspicuous geochemical feature on Mercury is the "high-Mg region" or HMR, as indicated on Figs. 2–4 and first identified by W15. A detailed investigation of the HMR and its possible origins was conducted by Frank et al. (2017), who used the maps reported here in their analysis. The higher-resolution maps presented here clarify compositional heterogeneity across the HMR that was hinted at in the maps of W15 (see Fig. 6). For example, a ~200-km-diameter region centered close to the crater Jobim (32.5°N, –67°E) has substantially lower Mg/Si and higher Al/Si than the HMR as a whole. This feature also corresponds to a topographic high, and thicker crust, than the rest of the HMR (Frank et al., 2017). Moreover, our maps reveal clear latitudinal variations in chemical composition across the HMR. In Fig. 6b we show profiles of Mg/Si, Al/Si, S/Si, and Ca/Si for six latitudinal bins across the HMR (the borders of these bins are indicated by red horizontal lines in Fig. 6a). Each profile is normalized to its average value to aid in comparison. We chose the widths of the bins to correspond approximately to the average spatial resolution within a bin, and the profiles exclude the low-Mg region near Jobim crater. The southernmost portion of the HMR (<8°N) is not included because the poorer spatial resolution causes too much overlap with material outside the HMR. We could not produce an Fe/Si profile in the same way because the areal coverage for Fe/Si is too sparse within the HMR. We find that the average Mg/Si ratio in the HMR is almost constant as a function of latitude, but is ~5% greater for latitudes >40°N. In contrast, the other three element ratios decrease substantially from south to north. In particular, the Al/Si ratio is some 20% lower in the northern part of the HMR than in the southern part. There is an uptick in S/Si and Ca/Si in the highest-latitude bin, but the coverage for these elements is not uniform in the northern portion of the HMR, so this may reflect in part sampling error. The mineralogical and petrologic implications of these variations deserve further study, but are beyond the scope of this work.

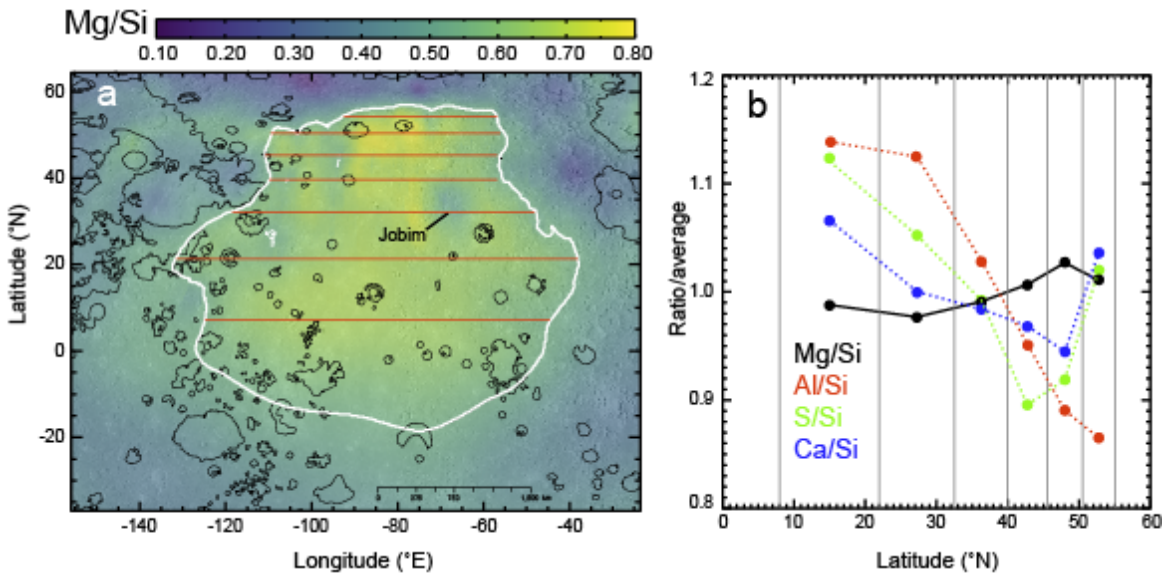

**Fig. 6.** (a) Smoothed map of Mg/Si, in cylindrical projection, for an area around Mercury's high-Mg region (HMR; outlined in white). A small area around Jobim crater has lower Mg/Si (and higher Al/Si) than the rest of the HMR. The horizontal red lines indicate boundaries of latitudinal

bins used to generate the profiles in (b). (b) Profiles of Mg/Si, Al/Si, S/Si, and Ca/Si as functions of latitude across the HMR (omitting the low-Mg region near Jobim crater); the ratios are normalized to their average values. Uncertainties in bin averages (from the standard error of the mean) are smaller than the plot symbols. The grey lines indicate bin boundaries.

*5.2.2. Northern smooth plains*

The NSP are a broad expanse of smooth plains, formed by flood volcanism at mid-to-high northern latitudes, that occupy some 7% of Mercury's surface (Head et al., 2011). The first global geochemical maps of Mercury showed that the NSP are not compositionally homogeneous, and W15 defined two geochemical terranes within them on the basis of their Mg/Si and Al/Si ratios: the "Low-Mg" (LM) NSP" and the "Intermediate-Mg" (IM) NSP having Mg/Si < 0.3 and ≈ 0.3–0.6, respectively. Our Mg/Si and Al/Si maps are shown over a monochrome Mercury Dual Imaging System (MDIS) mosaic in orthographic projection in Fig. 7a. Yellow-dashed lines indicate the approximate border between these two terranes, which is defined following the W15 delineation. The separation of the NSP into the IM-NSP and LM-NSP is also supported by the differences in the fluxes of thermal neutrons (Peplowski et al., 2015) and fast neutrons (Lawrence et al., 2017), as previously reported. Although the XRS maps show that the north pole exhibits some of the lowest Mg/Si and Al/Si ratios observed on Mercury, we note that the viewing geometry close to the pole is highly unfavorable for accurate XRF analysis (mainly because of the near-grazing solar incidence angles), and these values should thus be viewed with some caution. That being said, similar low Mg and Al abundances are seen at a few places farther south within the LM-NSP, and these data are not subject to this same problem.

Beyond a separation into two terranes, the new higher-resolution XRS maps reveal additional chemical heterogeneity within the NSP. For example, the area shown in Fig. 7b (a close-up view of the Mg/Si map for the northern area indicated by a solid-yellow polygon in Fig. 7a; note the different range of Mg/Si values for the color scale) has noticeably higher Mg/Si than the surrounding LM-NSP. This chemical anomaly is spatially well correlated with three ~100-km-diameter impact craters (Gaudí, Stieglitz, and an unnamed crater), suggesting that the chemical anomaly is caused by excavation of compositionally distinct material from beneath the flood basalts that comprise the LM-NSP.

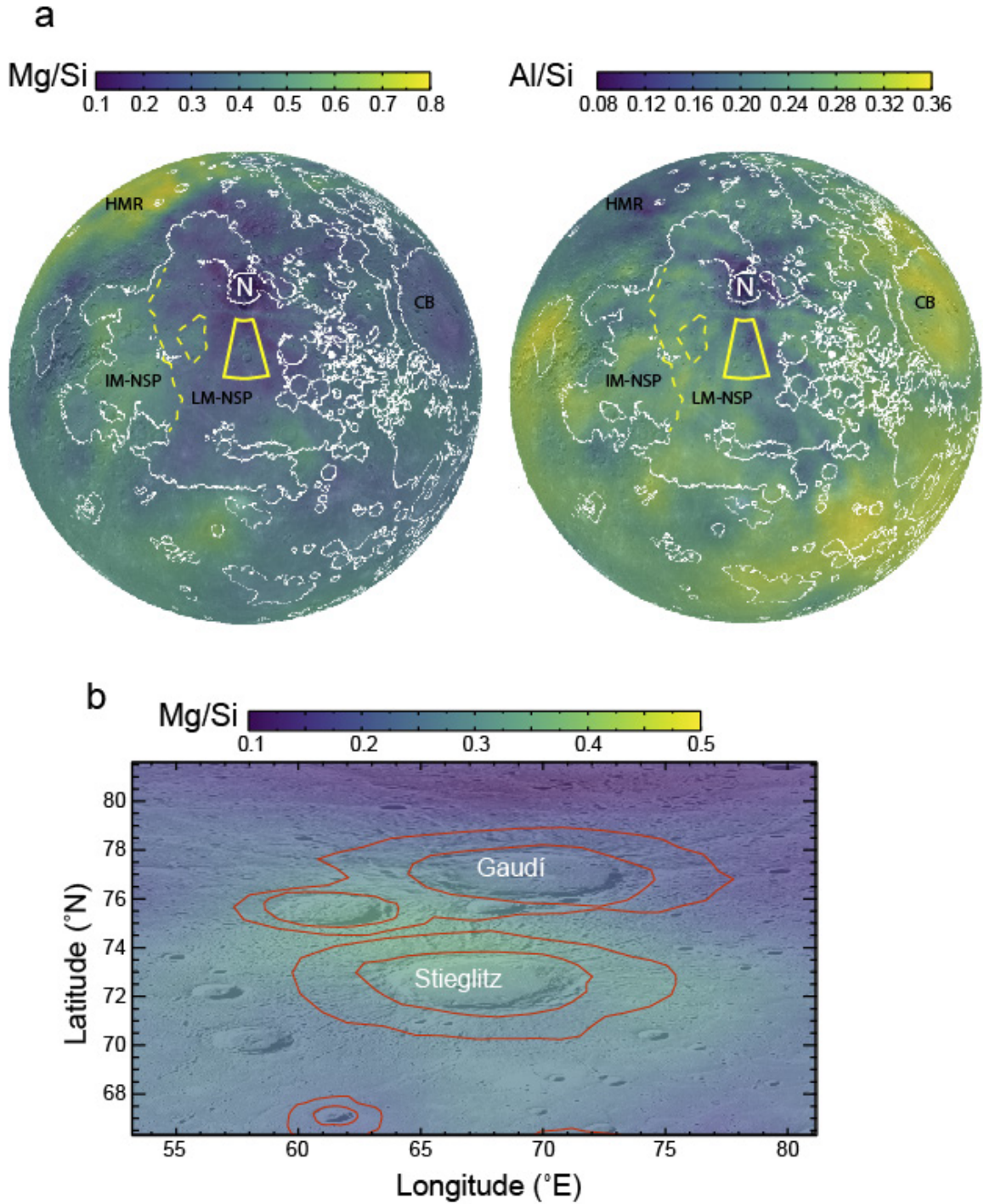

**Fig. 7.** (a) Smoothed maps of Mg/Si and Al/Si for Mercury's northern hemisphere, shown in orthographic projection centered on 60°N, 67°E. Smooth plains deposits (Denevi et al., 2013) are outlined in white. Several features are labeled: HMR = high-Mg region, IM/LM-NSP = intermediate-Mg/low-Mg northern smooth plains (the boundary between these two terranes is indicated by the dashed yellow line), N = north pole, CB = Caloris basin.

(b) Smoothed map of Mg/Si ratio, in cylindrical projection, for a region (marked with a yellow polygon in (a)) around Gaudí and Stieglitz craters. Note that the Mg/Si range for the color scale is different in (a) and (b). Red lines outline crater ejecta mapped by Prockter et al. (2016).

*5.2.3. Caloris basin*

The Caloris basin is the largest known impact basin on Mercury, with a diameter of 1550 km (Murchie et al., 2008). As seen in Fig. s 2–4, the XRS maps reveal that the high-reflectance interior plains of Caloris are Al-rich, with a high Al/Mg ratio and relatively low Mg/Si, S/Si, and Ca/Si ratios. The interior plains are also relatively low in K content, particularly in comparison with the NSP (Peplowski et al., 2012). The Al/Mg map of Caloris and its surroundings is shown in more detail in Fig. 8. Although, overall, the Al//Mg ratio of Caloris is among the highest observed on the planet, there are clearly resolved chemical differences across the basin. In particular, there are regions (e.g., those denoted by arrows in Fig. 8) with higher Mg/Si and lower Al/Si that correspond to impact craters that have excavated darker LRM material from depth and deposited it onto the brighter volcanic plains that dominate the basin (Ernst et al., 2015). Interestingly, our XRS maps do not indicate a distinct composition for the knobby and hummocky Odin-type plains (marked in Fig. 8) that are adjacent to Caloris compared with the other circum-Caloris smooth plains. Although it was originally postulated that the Odin-type plains were representative of Caloris ejecta deposits (e.g., Trask & Guest, 1975) because of their morphological resemblance to the ejecta facies of some lunar basins, their crater size–frequency distributions originally suggested that they were not emplaced as part of the Caloris impact event (Fassett et al., 2009), although the crater size–frequency data have recently been reinterpreted (Denevi et al., 2018). Whether the knobby and hummocky material of the Odin formation has an impact or volcanic (like the smooth plains adjacent to Caloris) origin therefore remains unclear. The similarity of the compositions of the Odin Formation and the circum-Caloris smooth plains, however, supports the interpretation that the knobby and hummocky material is indeed volcanic and that an unrecognized process gave rise to the Odin Formation's hummocky nature (e.g., Byrne et al., 2018; Denevi et al., 2013).

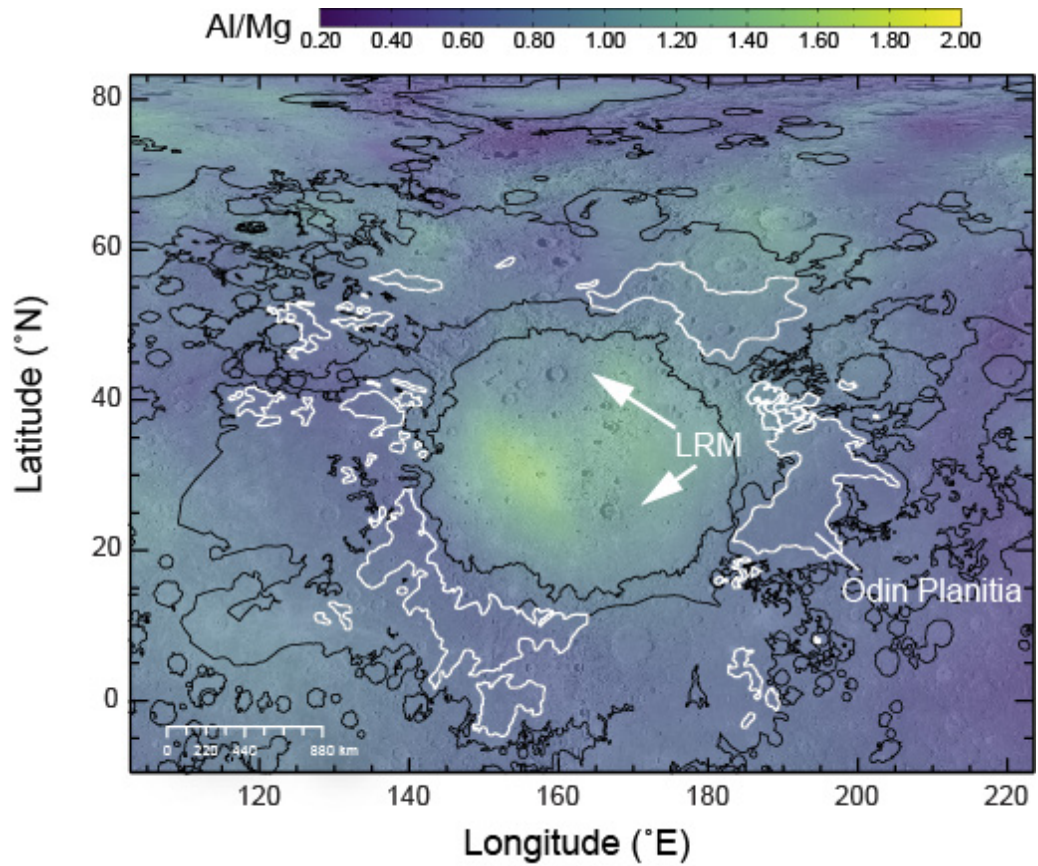

**Fig. 8**. Smoothed map of Al/Mg for the Caloris impact basin and surroundings, in cylindrical projection. Black lines denote smooth plains boundaries, and white lines outline the hummocky and knobby Odin-type plains (Odin Planitia) that surround the basin (Denevi et al., 2013). Arrows indicate areas within the basin that have lower Al/Mg as a result of the excavation of LRM from depth by the formation of impact craters.

*5.2.4. Lava flow boundaries*

Along its northwestern boundary, the HMR is in contact with a lobe of smooth plains deposits (Fig. 9) that are generally interpreted to be volcanic in origin (Denevi et al., 2018). Like other smooth plains, this unit has a lower density of superposed impact craters than the intercrater plains of which the HMR is mainly composed. The higher-resolution XRS measurements have refined chemical boundaries in this region and show a sharp contrast in composition, with the smooth plains having the low Mg/Si and intermediate Al/Si ratios that are characteristic of much of the NSP (e.g., Fig. 2). The contrast in crater densities suggests that this expanse of smooth plains, like the region surrounding it (which is mixed with ejecta from the Caloris basin to its west), was emplaced after the HMR. Even within this plains area, however, there is material with higher Mg/Si and lower Al/Si—most obviously near the Strindberg and Kōshō craters. Strindberg is a 187-km-diameter peak-ring crater (Baker et al., 2011) and its ejecta, as well as that of Kōshō, blankets the region of the smooth plains unit with material of high Mg/Si and low Al/Mg. Our high-resolution maps thus indicate that the material contained within the ejecta of Strindberg and Kōshō was derived, at least in part, from the excavation of a Mg-rich, sub-plains unit that was probably originally part of the HMR. The nearby LRM-rich Akutagawa crater is outside the smooth plains unit but is locally enriched in Mg/Si and depleted in Al/Si, as discussed in Section 5.1.

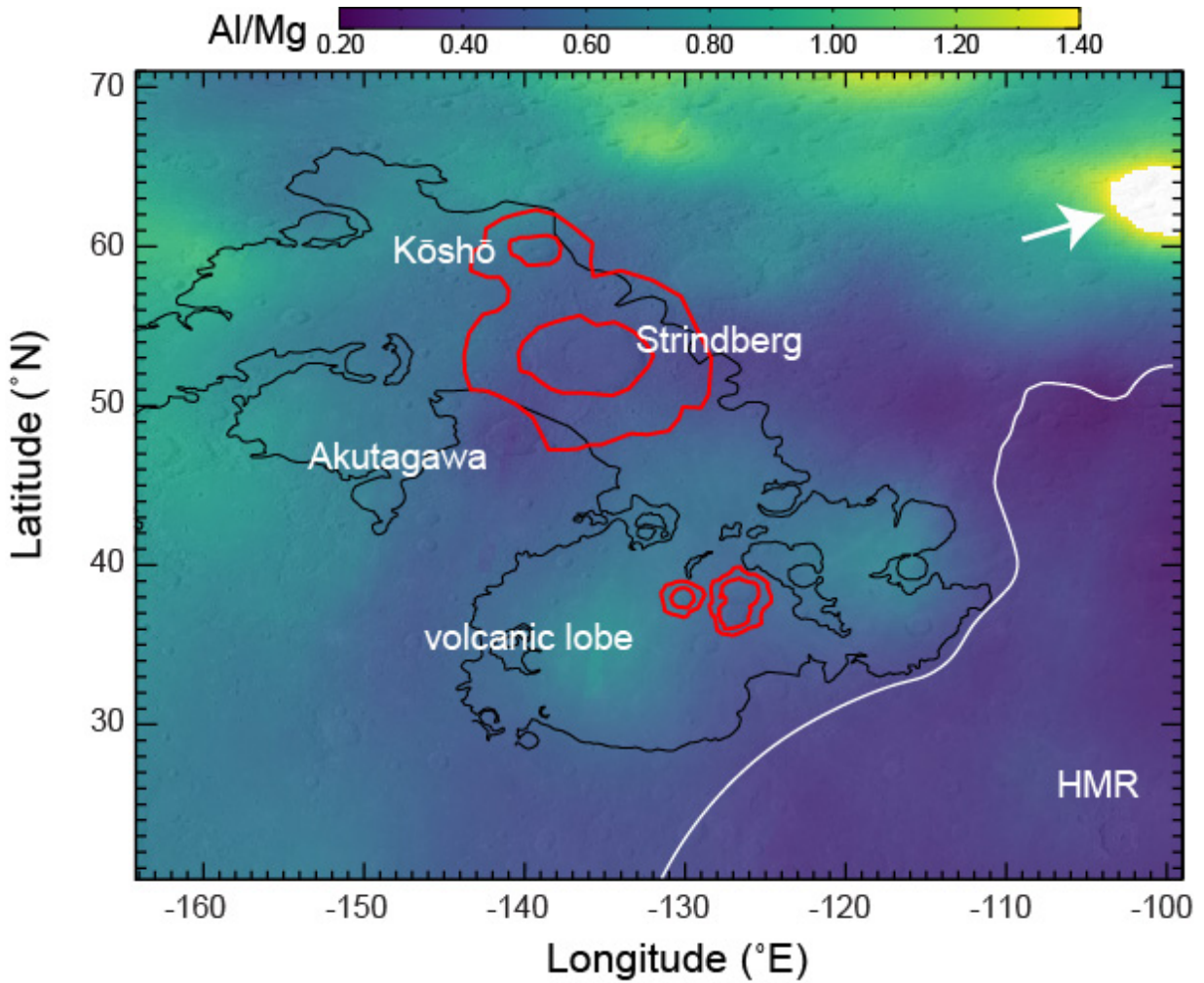

**Fig. 9.** Smoothed map of Al/Mg, in cylindrical projection, for a region northwest of Mercury's high-Mg region (HMR, white outline). Note the smaller data range, compared with Fig. 2e; values higher than 1.4 appear white. Black outlines denote boundaries of a region of smooth plains

(Denevi et al., 2013), including a volcanic lobe (so labeled) with distinctly higher Al/Mg than surrounding material; the lobe was likely emplaced later than its surroundings. Red outlines denote ejecta from selected impact impact craters as mapped by Prockter et al. (2016). The white arrow indicates a nearby location that exhibits the highest Al/Mg ratio, ~2, observed on the planet.

*5.2.5. High Al/Mg material*

Visible at the upper right of Fig. 9 (arrow) is a region, <100 km in diameter, with low Mg/Si (~0.13) and high Al/Si (~0.25), giving the highest Al/Mg ratio observed on Mercury. Unfortunately, we have no data for S, Ca, or Fe for this area. A closer look at this region (Fig. 10) shows the Al/Mg map for the area, together with monochrome and enhanced-color mosaics derived from MDIS images (Denevi et al., 2018; Murchie et al., 2015). The enhanced-color map is a red-green-blue (RGB) image derived from a principal component (PC) analysis of eight-color maps, where the red channel corresponds to the second PC, green to the first PC, and blue to 430-nm/1000-nm spectral reflectance ratio. In this representation, the spectral end-members of high-reflectance red plains and LRM appear as bright orange and dark blue, respectively, and fresh material that is relatively unaffected by space weathering appears bright and less red. Figure 10 illustrates that the peak of the high Al/Mg region is concentrated just outside the NSP, with which it shares a similar Al and Mg composition, and that it lies south of the 70-km-diameter Gaugin crater. The high Al/Mg region also coincides with an area that appears brighter and more cyan in color in the enhanced-color map, which indicates relatively fresh (recently excavated) material. We note that another spot of high Al/Mg just outside the NSP is also visible in Fig. 10, at ~60°N, –70°E. This anomaly is aligned with another bright impact crater, although this crater does not exhibit the cyan color indicative of freshly excavated material in Fig. 10c (another crater just to the west, however, does display the cyan signature).

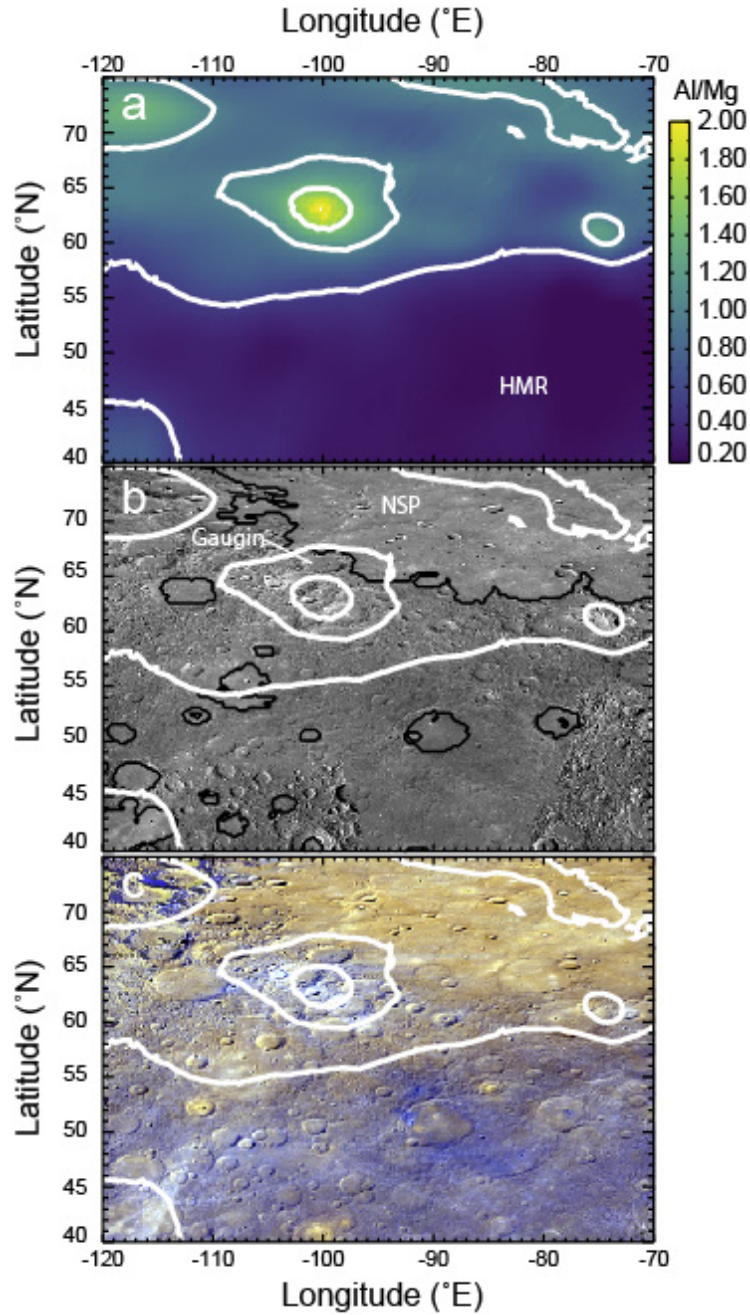

**Fig. 10.** (a) Smoothed map of Al/Mg, (b) monochrome MDIS mosaic, and (c) enhanced-color MDIS mosaic (see text for explanation) for a location north of the high-Mg region (HMR) and close to Gaugin crater that includes a region with the highest Al/Mg ratio observed on Mercury. The white curves are contours of Al/Mg ratio, and black lines denote smooth plains boundaries (Denevi et al., 2013).

## 6. Conclusions

The MESSENGER XRS allowed measurements of Mg/Si, Al/Si, S/Si, Ca/Si, and Fe/Si weight ratios to be made across Mercury's surface. We have used XRS data collected throughout the entire >4-year orbital mission of MESSENGER to generate maps of these ratios, along with associated uncertainty and spatial resolution maps. We have described the procedures for data selection and map generation, including smoothing and the updated phase-angle correction for the Fe/Si data. The smoothed versions of these maps are publicly available in NASA's PDS, and both unsmoothed and smoothed versions are provided as Supplementary data for this paper.

Spatial coverage for each of the XRS element-ratio maps varies strongly. That is, the Mg/Si and Al/Si maps are based on tens of thousands of measurements collected under a range of solar conditions and have 100% global coverage, whereas the Fe/Si map is based on a few hundred measurements acquired during the largest solar flares and has only 40% areal coverage of the northern hemisphere and only 18% coverage northward of 30°N. The S/Si and Ca/Si maps are also based on flare measurements, but they have greater areal coverage than the Fe/Si map.

The spatial resolution of XRS measurements was highly variable because of MESSENGER's eccentric near-polar orbit and time-dependent periapsis altitude. The procedure we developed to generate maps from such data results in maps where the resolution can vary from pixel to pixel. This situation differs from that for typical mapped planetary data and means that each pixel value should be thought of as an average over a region of a given size surrounding it. Sharp changes in element ratio between adjacent pixels most likely reflect changes in resolution rather than in composition. Users of these maps are thus cautioned to consider the spatial resolution maps along with the element-ratio maps themselves when interpreting smaller-scale features on the maps.

We have provided in this paper several examples of how the XRS maps can be used to investigate elemental variations in the context of prominent geological features on Mercury, with spatial scales ranging from single ~100-km-diameter craters to large impact basins. Overall, the XRS maps presented here support the presence of previously reported large-scale geochemical terranes on Mercury. The smaller scale features within those terranes seem generally to correlate with impact craters and are thus compositional overprints caused by the excavation of chemically distinct material from beneath the surface. On the basis of the examples presented here, we expect that these maps will continue to contribute to studies of Mercury's origin and geological history for many years to come.

**Acknowledgements**

We thank the entire MESSENGER team for the development, launch, cruise, orbit insertion, and orbital operations of the MESSENGER spacecraft. Comments by Patrick Peplowski and an anonymous referee improved this paper. We acknowledge Stéfan van der Walt and Nathaniel Smith for development of the "viridis" colormap used here. This work was supported by the NASA Discovery Program under contract NAS5–97271 to The Johns Hopkins University Applied Physics Laboratory and NASW-00002 to the Carnegie Institution of Washington.

Table 1. Number of XRS spectra used for generation of element-ratio maps.

|  |  | Mg/Si | Al/Si | S/Si | Ca/Si | Fe/Si |
|---|---|---|---|---|---|---|
| Solar flares | | 1,065 | 1,065 | 1,386 | 1,419 | 262 |
| Quiet Sun | >100 km | 33,954 | 33,954 | | | |
| | 50–100 km | 10,372 | 10,372 | | | |
| | <50 km | 2,337 | 2,337 | | | |
| | Total | 46,663 | 46,663 | | | |
| | Total (binned) | 22,801 | 22,801 | | | |

Supplementary Information for Nittler et al.: **Global Major-Element Maps of Mercury from Four Years of MESSENGER X-Ray Spectrometer Observations**

**Supplemental Figures**

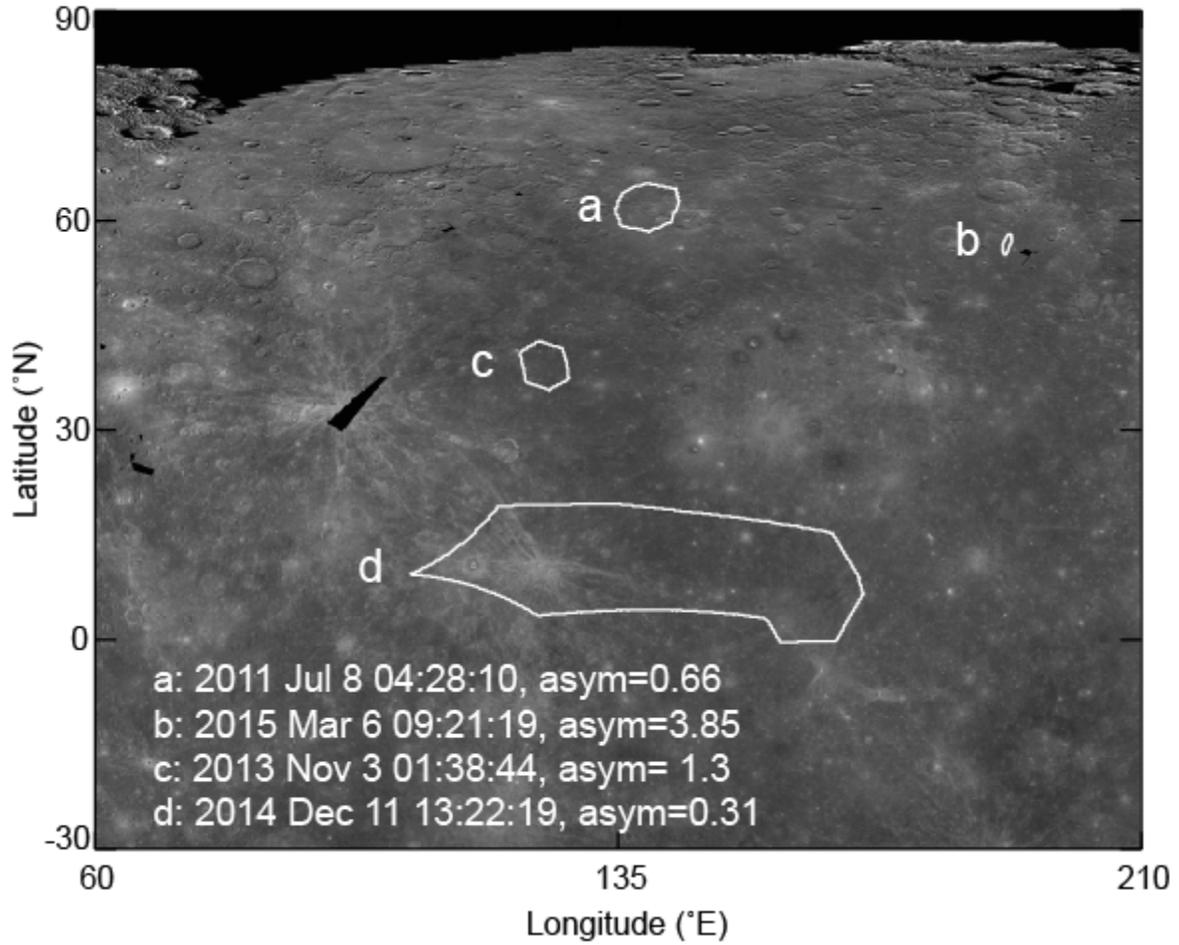

**Fig. S1**. Example XRS footprints, with their calculated asymmetry ("asym") values (defined as the maximal north–south extent of the footprint, in kilometers, divided by the maximal east–west extent).

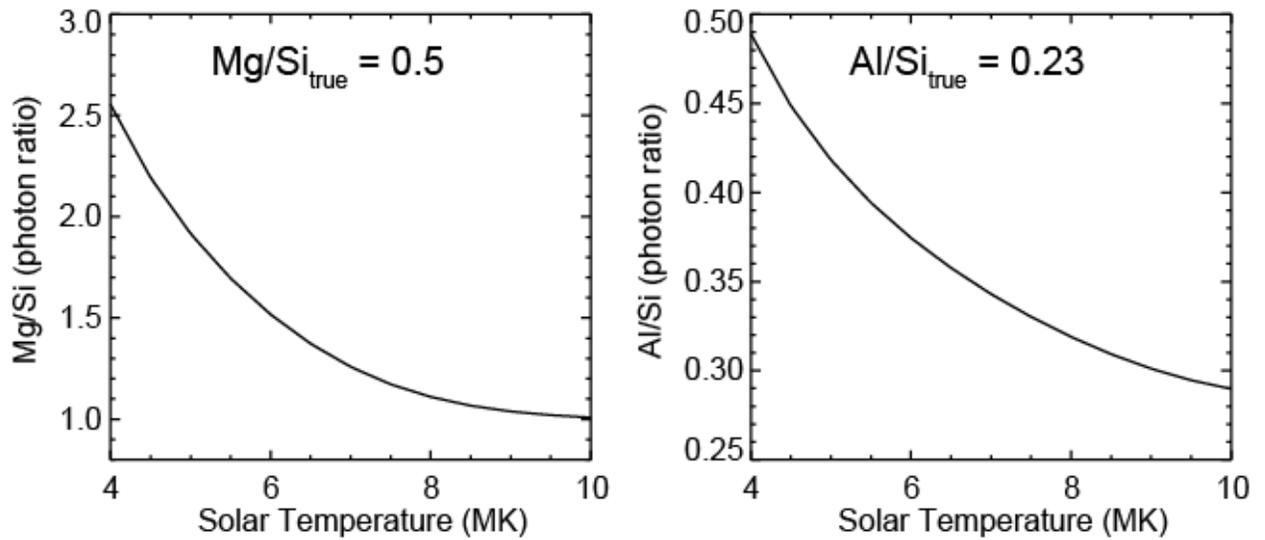

**Fig. S2**. Calibration curves relating predicted Mg/Si and Al/Si photon ratios as a function of solar coronal temperature. Curves scale linearly for different true values of Mg/Si and Al/Si. These curves are used to convert quiet-Sun XRF fluxes into elemental ratios. They are empirically corrected to match solar flare data as described by W15.

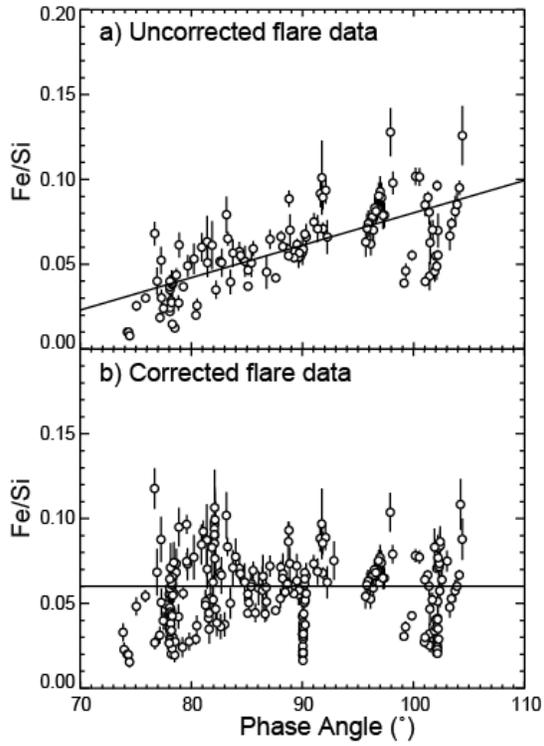

**Fig. S3.** (a) Measured Fe/Si ratios as a function of phase (Sun–planet–XRS) angle for 117 flare measurements with large footprints in Mercury's southern hemisphere. The linear fit (solid line) yields Fe/Si = 0.00191 $\phi$ – 0.11068, where $\phi$ is the phase angle in degrees. (b) Corrected Fe/Si data (see main text).

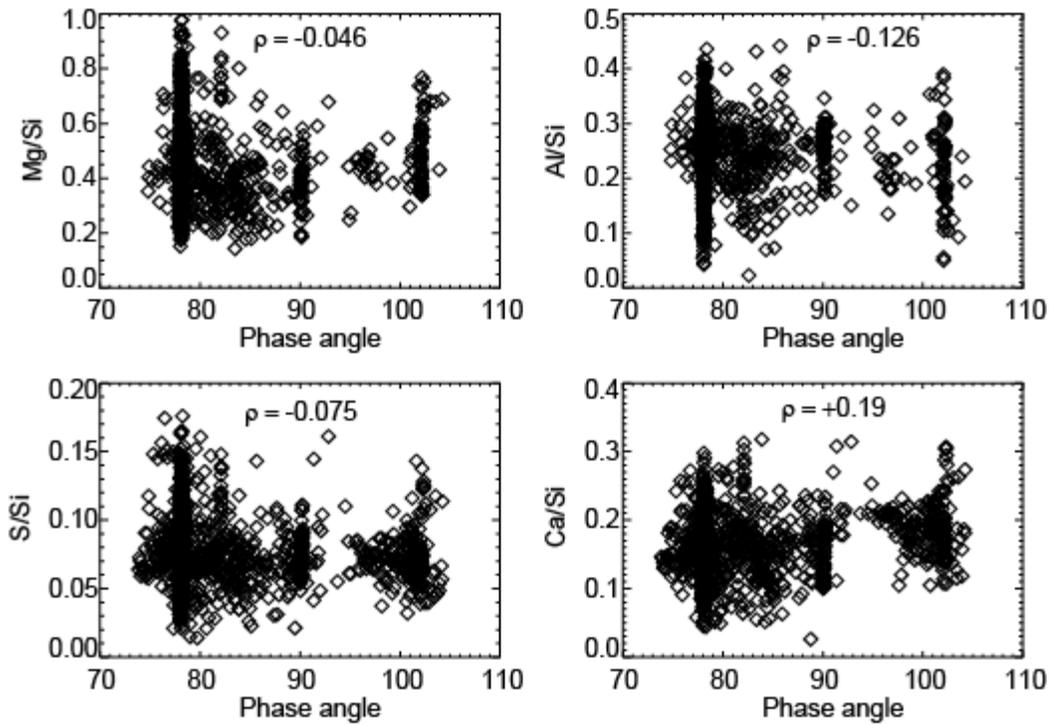

**Fig. S4.** Measured element ratios as a function of phase angle (degrees) for XRS flare measurements used to generate element maps. Pearson's correlation coefficients are indicated on panels. There are slight negative correlations between Mg/Si, Al/Si, and S/Si and phase angle, and a slight positive correlation for Ca/Si. However, the changes in average ratios over the full range of phase angle are much smaller than the scatter of the data for any given angle, and thus there is no significant indication of a phase-angle effect for these element ratios.

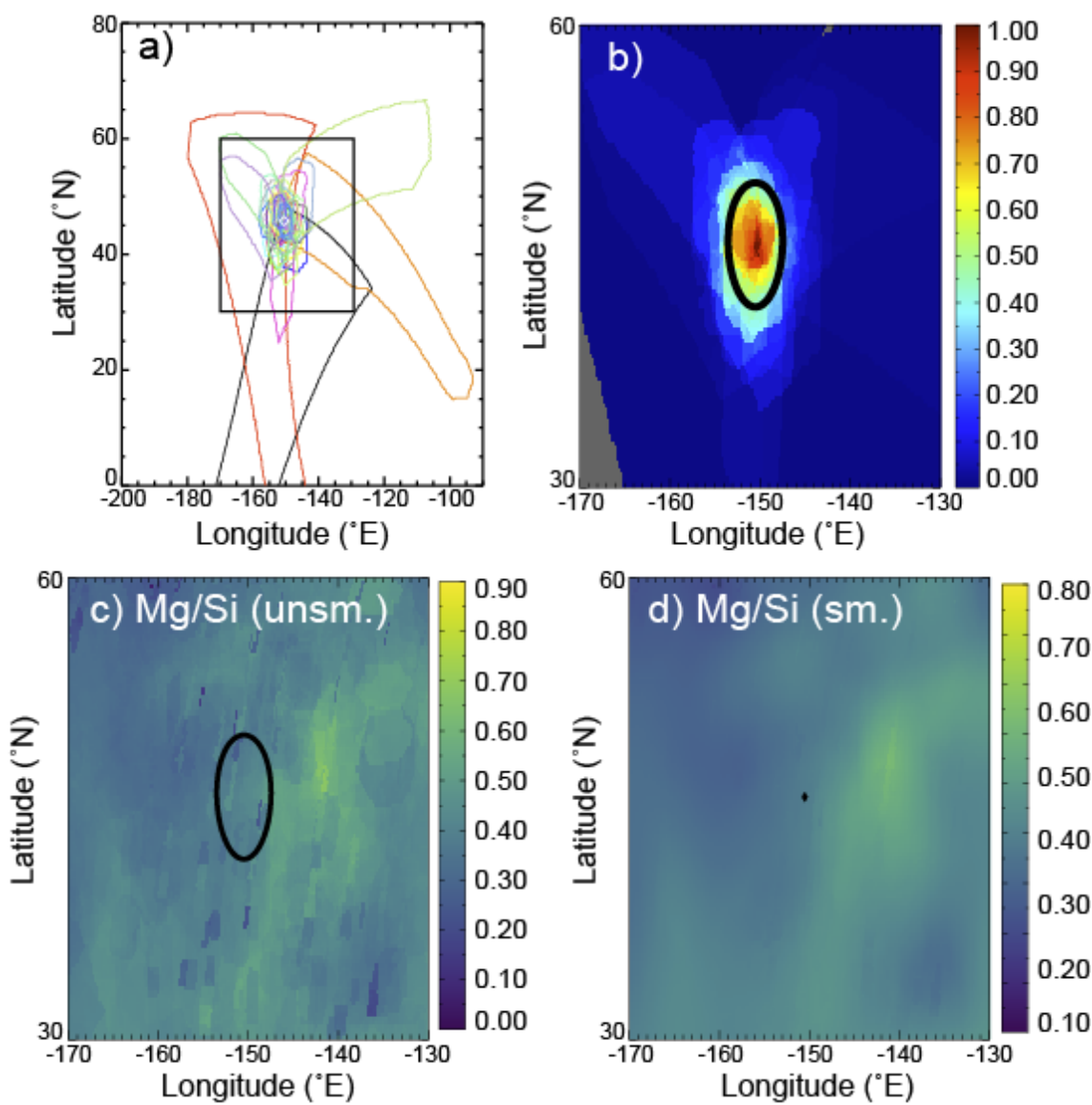

**Fig. S5.** XRS footprint averaging and smoothing. (a) Outlines of all XRS footprints that overlap the map pixel corresponding to 45.75°N, -150.5°E. The rectangle indicates the location of panels (b–d). (b) Weighted average of footprints from (a), weighted by footprint area and Mg/Si uncertainty as discussed in the text and normalized to the maximum value. Most of the contribution to the map value comes from a central region where values are highest. (c) Unsmoothed and (d) smoothed Mg/Si map of the same area. The ellipse on (b) and (c) indicates the smoothing ellipse defined by the weighted asymmetry parameter (Fig. S1) over all the footprints from (a) and the weighted average footprint area, divided by 2. This ellipse is a good approximation to the >50% threshold of the weighted footprint in (b). The final pixel value in the smoothed map (black point

in (d)) is the average of all pixels within the ellipse in (c). This process is repeated for all pixels to yield the complete smoothed map.

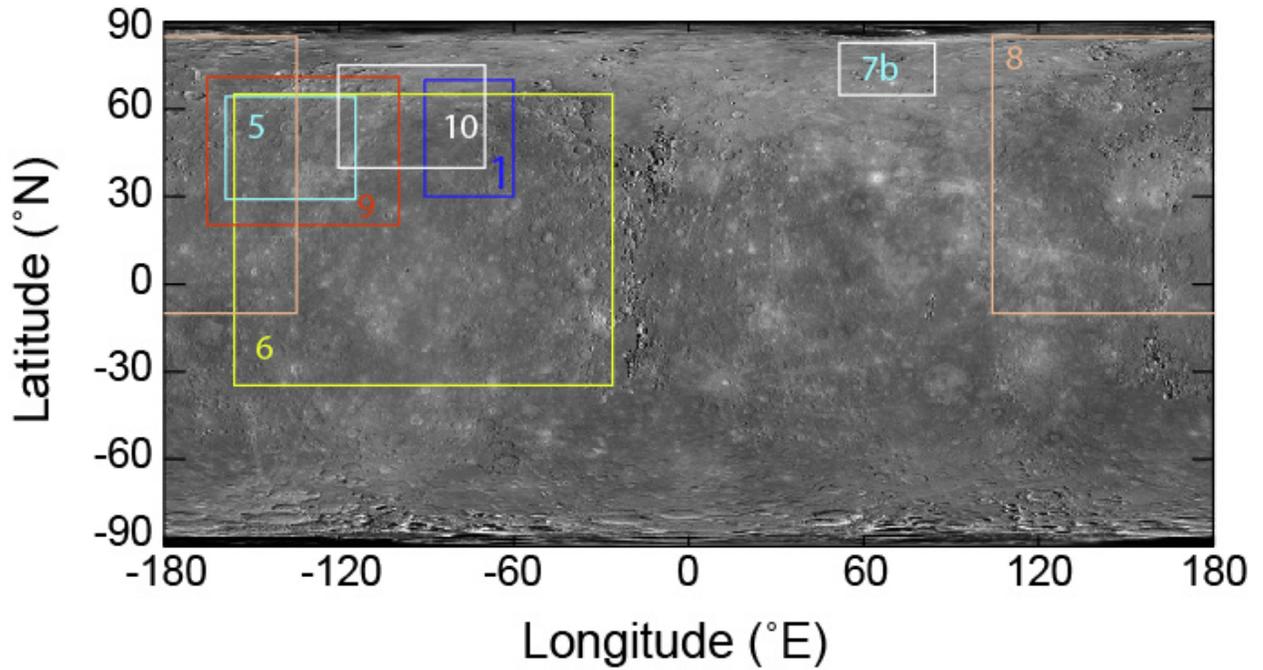

**Fig. S6.** Overview of the locations of the specific regions discussed in this paper and shown in other Figures. Rectangles indicate the locations of figures in the main text and their corresponding figure numbers, overlain on an MDIS monochrome image mosaic in cylindrical projection.